# Lattice Boltzmann Method Simulation of 3-D Natural Convection with Double MRT Model

Zheng Li[a,b], Mo Yang[a] and Yuwen Zhang[b,1]

[a] College of Energy and Power Engineering, University of Shanghai for Science and Technology, Shanghai 200093, China
[b] Department of Mechanical and Aerospace Engineering, University of Missouri, Columbia, MO 65211, USA

**ABSTRACT**

Multiple-relaxation-time model (MRT) has more advantages than the many others approaches in the Lattice Boltzmann Method (LBM). Three-dimensional double MRT model is proposed for the first time for fluid flow and heat transfer simulation. Three types of cubic natural convection problems are solved with proposed method at various Rayleigh numbers. Two opposite vertical walls on the left and right are kept at different temperatures for all three types, while the remained four walls are either adiabatic or have linear temperature variations. For the first two types of cubic natural convections that four walls are either adiabatic or vary linearly, the present results agreed very well with the benchmark solutions or experimental results in the literature. For the third type of cubic natural convection, the front and back surfaces has linearly variable temperature while the bottom and top surface are adiabatic. The results from the third type exhibited more general three-dimensional characters.

**NOMENCLATURE**
$c$    lattice speed
$c_p$   specific heat (J/kgK)
$c_s$   sound speed
$e_i$   particle speed
$f_i$   density distribution
$F_i$   body force
$g$    gravitational acceleration $(m/s^2)$
$G$   effective gravitational acceleration $(m/s^2)$
$k$    thermal conductivity (W/m k)
$M$   transform matrix for density distribution
$m_i$   moment function for density distribution
$Ma$   Mach number
$N$   transform matrix for density distribution
$n_i$   moment function for energy distribution

---

[1] Corresponding author. Email: zhangyu@missouri.edu. Tel: 001-573-884-6936. Fax: 001-573-884-5090

$p$ pressure (Pa)
$P$ non-dimensional pressure
$Pr$ Prandtl number
$Q$ collision matrix for energy distribution
$Ra$ Rayleigh number
$S$ collision matrix for density distribution
$t$ time (s)
$T$ temperature $(K)$
$U$ non-dimensional velocity in x-direction
$v$ velocity in y-direction (m/s)
$V$ non-dimensional velocity in y-direction
$w$ velocity in y-direction (m/s)
$W$ non-dimensional velocity in y-direction
$V$ velocity
$\alpha$ thermal diffusivity $(m^2/s)$
$\beta$ thermal expansion (K$^{-1}$)
$\Delta t$ time step (s)
$\theta$ non-dimensional temperature
$\mu$ viscosity (Kg/ms)
$\rho$ Density (kg/m$^3$)
$\nu$ kinematic viscosity $(m^2/s)$
$\omega_i$ value factor for velocity
$\omega_{T,i}$ value factor for temperature

# 1. Introduction

Lattice Boltzmann method (LBM) has become an increasingly popular numerical method in the last three decades. It has been used to simulate various hydrodynamic systems, such as incompressible fluid flow [1, 2], porous media flow [3], and melting problem [4, 5]. Different from traditional computational fluid dynamics (CFD) approaches, LBM is based on the discrete Boltzmann equation in statistical physics and it has two basic steps: collision step and streaming steps [6]. LBM has several models, which differs from each other with the method to handle the collision step. The most popular one is the Lattice Bhatnagar-Gross-Krook (LBGK) [7, 8] that simplifies collision step with a single relaxation time term. Some shortcomings of LBGK are also apparent that Prandtl number must be unity when the model is applied to thermal fluids and it suffers numerical instability [9, 10]. To overcome this limitation, entropy LBM (ELBM) [11, 12], two-relaxation-time model (TRT) [13, 14] and multiple relaxation time model (MRT) [15-17] have been proposed. Luo et al. [18] compared various LBM models and concluded that MRT was preferred due to its advantages in accuracy and numerical stability. This article uses MRT model to simplify the collision step.

Lattice Boltzmann method was only valid to solve fluid flow when it was proposed [7]. Fluid flow involved with heat transfer problem is important due to its numerous applications in industry fields. Multispeed approach (MS), hybrid method and double distribution functions (DDF) are three common thermal LBM models. MS obtains the temperature field by adding more discrete velocities to density distribution [19]; it is limited by numerical instability and narrow range of temperature variation [20]. In hybrid method, the velocity field is solved with LBM while temperature field is obtained using other numerical methods, such as finite volume method (FVM) and Monte Carlo method. Li et al. [21, 22] solved natural convection and melting problems with a hybrid LBM and FVM method. Hybrid LBM and MCM method was proved to be suitable for convection simulation [23]. Besides density distribution, DDF includes an addition distribution to analyze the temperature field. Huber et al. [24] proposed a DDF model for coupled diffusions based on MRT. This article employs DDF with MRT to simulate the heat transfer process.

Natural convection plays an important role in many industry fields. Numerous two-dimensional results were reported to discuss natural convection effect in various processes [4, 5, 20-23, 25-29]. The research on three-dimensional cases are scant although they are more general. Leong et al. [30] provided experimental Nusselt numbers for a cubical-cavity benchmark problem in three-dimensional natural convection. Tric et al. [31] solved three-dimensional natural convention at different Rayleigh numbers numerically. Wakashima and Saitoh [32] obtained Benchmark solutions for natural convection in a cubic cavity using the high-order time-space method. Salat et al. [33] did experimental and numerical investigation of turbulent natural convection in a large air-filled cavity. This problem was also investigated using LBM [34]. Azwadi and Syahrullall [35] employed a double LBGK model to simulate natural convection in a cubic cavity. LBM models in DDF can be different from each other. MRT and LBGK were applied to solve velocity and temperature fields respectively in a cubic cavity mix convection flow [36, 37]. This article proposed a double MRT model, which has not been reported by our knowledge, for three-dimensional fluid flow and heat transfer simulation. The objective of this paper is to discuss three-dimensional natural convections with proposed double MRT model.

## 2. Natural Convection in a Cubic Cavity

Physical model of natural convection in a cubic cavity is shown in Fig. 1. The cubic cavity with an edge length of H is filled with working fluid of air. The Prandtl number is fixed at 0.71. Boussinesq assumption is employed. Then the governing equations are

$$\frac{\partial u}{\partial x}+\frac{\partial v}{\partial y}+\frac{\partial w}{\partial z}=0 \tag{1}$$

$$\rho\left(\frac{\partial u}{\partial t}+u\frac{\partial u}{\partial x}+v\frac{\partial u}{\partial y}+w\frac{\partial u}{\partial z}\right)=-\frac{\partial p}{\partial x}+\mu\left(\frac{\partial^2 u}{\partial x^2}+\frac{\partial^2 u}{\partial y^2}+\frac{\partial^2 u}{\partial z^2}\right) \tag{2}$$

$$\rho\left(\frac{\partial v}{\partial t}+u\frac{\partial v}{\partial x}+v\frac{\partial v}{\partial y}+w\frac{\partial v}{\partial z}\right)=-\frac{\partial p}{\partial y}+\mu\left(\frac{\partial^2 v}{\partial x^2}+\frac{\partial^2 v}{\partial y^2}+\frac{\partial^2 v}{\partial z^2}\right) \tag{3}$$

$$\rho\left(\frac{\partial w}{\partial t}+u\frac{\partial w}{\partial x}+v\frac{\partial w}{\partial y}+w\frac{\partial w}{\partial z}\right)=-\frac{\partial p}{\partial z}+\mu\left(\frac{\partial^2 w}{\partial x^2}+\frac{\partial^2 w}{\partial y^2}+\frac{\partial^2 w}{\partial z^2}\right)+\rho g \beta (T-T_l) \tag{4}$$

$$\rho\left(\frac{\partial T}{\partial t}+u\frac{\partial T}{\partial x}+v\frac{\partial T}{\partial y}+w\frac{\partial T}{\partial z}\right)=k\left(\frac{\partial^2 T}{\partial x^2}+\frac{\partial^2 T}{\partial y^2}+\frac{\partial^2 T}{\partial z^2}\right) \tag{5}$$

Three types of natural convections are in consideration in this article. They differ from each other with the thermal boundary settings. In all these cubic cavity natural convections, the vertical walls ($y = 0$ and $y = H$) are kept at $T_h$ and $T_c$ respectively. $u$, $v$ and $w$ are velocities in the x-, y- and z-directions. Non-slip boundary conditions are employed.

$$y = 0,\ T = T_h \tag{6}$$

$$y = H,\ T = T_c \tag{7}$$

$$u = v = w = 0 \text{ for all boundaries} \tag{8}$$

In the first type cubic cavity natural convection, the remained four walls are all adiabatic. Natural convection with this setting is solved as benchmark problem in Ref. [31-34]. Leong et al. [30] argued that this setting was not physically-realizable because adiabatic boundary condition was hard to be reached.

The second type of cubic cavity natural convection is a physically-realizable benchmark problem. Its remained four walls have a linear temperature variation from the cold surface to hot surface. Experimental results are reported in Refs. [30, 38]. In this article, we use:

$$\begin{cases} x = 0,\ T = T_h + (T_c - T_h) \cdot y/H \\ x = H,\ T = T_h + (T_c - T_h) \cdot y/H \\ z = 0,\ T = T_h + (T_c - T_h) \cdot y/H \\ z = H,\ T = T_h + (T_c - T_h) \cdot y/H \end{cases} \tag{9}$$

The following boundary conditions are applied to the third type cubic natural convection.

$$\begin{cases} x = 0,\ T = T_h + (T_c - T_h) \cdot y/H \\ x = H,\ T = T_h + (T_c - T_h) \cdot y/H \\ z = 0,\ \partial T/\partial z = 0 \\ z = H,\ \partial T/\partial z = 0 \end{cases} \tag{10}$$

In the remained parts of this article, these three types of problems are referred to as Type 1, Type 2 and Type 3.

## 3. Lattice Boltzmann method

Lattice Boltzmann method has been employed to solve many fluid flow and heat transfer problems [20, 22, 37]. MS, hybrid method and DDF are the common thermal LBM models. The DDF in LBM will be used to solve velocity and temperature fields, respectively.

Streaming and collision are the basic processes in LBM. Many LBM models exist and they differ from each other by the ways to simplify the collision term. MRT is selected due to its advantages in accuracy and numerical stability [18]. To the best of the authors' knowledge, double MRT model for three-dimensional fluid flow and heat transfer simulation has not been reported by now. In this article, 3-D double MRT model is proposed and three types of natural convections are solved with it for validation.

## 3.1 D3Q19-MRT-LBM Model for Fluid Flow

Lattice Boltzmann equation can describe the statistical behavior of a fluid flow.

$$f(\mathbf{r}+\mathbf{e}\Delta t, t+\Delta t) - f(\mathbf{r},t) = \Omega + F \quad (11)$$

where $f$ is the density distribution, $\Delta t$ is the time step, $\Omega$ is the collision term and $F$ is the body force. To simplify this equation, it is assumed that each computing nodes has 19 directions as shown in Fig. 2 and these velocities are given by:

$$\mathbf{e}_i = c \begin{bmatrix} 0 & 1 & -1 & 0 & 0 & 0 & 0 & 1 & -1 & 1 & -1 & 1 & -1 & 1 & -1 & 0 & 0 & 0 & 0 \\ 0 & 0 & 0 & 1 & -1 & 0 & 0 & 1 & 1 & -1 & -1 & 0 & 0 & 0 & 0 & 1 & -1 & 1 & -1 \\ 0 & 0 & 0 & 0 & 0 & 1 & -1 & 0 & 0 & 0 & 0 & 1 & 1 & -1 & -1 & 1 & 1 & -1 & -1 \end{bmatrix} \quad (12)$$

where $c$ is the lattice speed and it relates to the sound speed $c_s$ as:

$$c^2 = 3c_s^2 \quad (13)$$

In this D3Q19 model, Eq. (11) can be expressed as:

$$f_i(\mathbf{r}+\mathbf{e}_i\Delta t, t+\Delta t) - f_i(\mathbf{r},t) = \Omega_i + F_i, \quad i = 1, 2, \ldots 19 \quad (14)$$

where the force term in the equation can be obtained as:

$$F_i = \Delta t \mathbf{G} \cdot \frac{(\mathbf{e}_i - \mathbf{V})}{p} f_i^{eq}(\mathbf{r},t) \quad (15)$$

where $\mathbf{G}$ is the effective gravitational force:

$$\mathbf{G} = -\beta(T - T_l)\mathbf{g} \quad (16)$$

The equilibrium distribution function $f_i^{eq}(\mathbf{r},t)$ is expressed as:

$$f_i^{eq}(\mathbf{r},t) = \rho \omega_i \left[ 1 + \frac{\mathbf{e}_i \cdot \mathbf{V}}{c_s^2} + \frac{(\mathbf{e}_i \cdot \mathbf{V})^2}{2c_s^4} - \frac{\mathbf{V} \cdot \mathbf{V}}{2c_s^2} \right], \quad i = 1, 2, \ldots 19 \quad (17)$$

The density weighting factors $\omega_i$ are:

$$\omega_i = \begin{cases} 1/3, & i = 1 \\ 1/18, & i = 2, 3, \cdots, 7 \\ 1/36, & i = 8, 9, \cdots, 19 \end{cases} \quad (18)$$

To satisfy the continuum and momentum conservations, the collision term in MRT is:

$$\Omega_i = -M^{-1} \cdot S \cdot \left[ m_i(\mathbf{r},t) - m_i^{eq}(\mathbf{r},t) \right], \quad i = 1, 2, \ldots 19 \quad (19)$$

where $m_i(\mathbf{r},t)$ and $m_i^{eq}(\mathbf{r},t)$ are moments and their equilibrium functions; $M$ and $S$

are the transform matrix and collision matrix respectively. d'Humieres et al. [10] introduced the detailed parameter settings in D3Q19-MRT model.

Macroscopic parameters are relate to the density distributions as:

$$\begin{cases} \rho = \sum_{i=1}^{19} f_i(\mathbf{r},t), \ \rho V = \sum_{i=1}^{19} f_i(\mathbf{r},t) \\ \mathbf{m} = M \cdot [f_1, \ f_2, \ \cdots \ f_{19}]^T \end{cases} \quad (20)$$

In this model, transform matrix $M$ is:

$$M = \begin{bmatrix}
1 & 1 & 1 & 1 & 1 & 1 & 1 & 1 & 1 & 1 & 1 & 1 & 1 & 1 & 1 & 1 & 1 & 1 & 1 \\
-30 & -11 & -11 & -11 & -11 & -11 & -11 & 8 & 8 & 8 & 8 & 8 & 8 & 8 & 8 & 8 & 8 & 8 & 8 \\
12 & -4 & -4 & -4 & -4 & -4 & -4 & 1 & 1 & 1 & 1 & 1 & 1 & 1 & 1 & 1 & 1 & 1 & 1 \\
0 & 1 & -1 & 0 & 0 & 0 & 0 & 1 & -1 & 1 & -1 & 1 & -1 & 1 & -1 & 0 & 0 & 0 & 0 \\
0 & -4 & 4 & 0 & 0 & 0 & 0 & 1 & -1 & 1 & -1 & 1 & -1 & 1 & -1 & 0 & 0 & 0 & 0 \\
0 & 0 & 0 & 1 & -1 & 0 & 0 & 1 & 1 & -1 & -1 & 0 & 0 & 0 & 0 & 1 & -1 & 1 & -1 \\
0 & 0 & 0 & -4 & 4 & 0 & 0 & 1 & 1 & -1 & -1 & 0 & 0 & 0 & 0 & 1 & -1 & 1 & -1 \\
0 & 0 & 0 & 0 & 0 & 1 & -1 & 0 & 0 & 0 & 0 & 1 & 1 & -1 & -1 & 1 & 1 & -1 & -1 \\
0 & 0 & 0 & 0 & 0 & -4 & 4 & 0 & 0 & 0 & 0 & 1 & 1 & -1 & -1 & 1 & 1 & -1 & -1 \\
0 & 2 & 2 & -1 & -1 & -1 & -1 & 1 & 1 & 1 & 1 & 1 & 1 & 1 & 1 & -2 & -2 & -2 & -2 \\
0 & -4 & -4 & 2 & 2 & 2 & 2 & 1 & 1 & 1 & 1 & 1 & 1 & 1 & 1 & -2 & -2 & -2 & -2 \\
0 & 0 & 0 & 1 & 1 & -1 & -1 & 1 & 1 & 1 & 1 & -1 & -1 & -1 & -1 & 0 & 0 & 0 & 0 \\
0 & 0 & 0 & -2 & -2 & 2 & 2 & 1 & 1 & 1 & 1 & -1 & -1 & -1 & -1 & 0 & 0 & 0 & 0 \\
0 & 0 & 0 & 0 & 0 & 0 & 0 & 1 & -1 & 1 & -1 & 0 & 0 & 0 & 0 & 0 & 0 & 0 & 0 \\
0 & 0 & 0 & 0 & 0 & 0 & 0 & 0 & 0 & 0 & 0 & 0 & 0 & 0 & 0 & 1 & -1 & -1 & 1 \\
0 & 0 & 0 & 0 & 0 & 0 & 0 & 0 & 0 & 0 & 0 & 1 & -1 & -1 & 1 & 0 & 0 & 0 & 0 \\
0 & 0 & 0 & 0 & 0 & 0 & 0 & 1 & -1 & 1 & -1 & -1 & 1 & -1 & 1 & 0 & 0 & 0 & 0 \\
0 & 0 & 0 & 0 & 0 & 0 & 0 & -1 & -1 & 1 & 1 & 0 & 0 & 0 & 0 & 1 & -1 & 1 & -1 \\
0 & 0 & 0 & 0 & 0 & 0 & 0 & 0 & 0 & 0 & 0 & 1 & 1 & -1 & -1 & -1 & -1 & 1 & 1
\end{bmatrix} \quad (21)$$

The corresponding macroscopic moments are:

$$\mathbf{m} = (\rho, \ e, \ \varepsilon, \ j_x, \ q_x, \ j_y, \ q_y, \ j_z, \ q_z, \ 3p_{xx}, \ 3\pi_{xx}, \ p_{ww}, \ \pi_{ww}, \ p_{xy}, \ p_{yz}, \ p_{xz}, \ m_x, \ m_y, \ m_z)^T \quad (22)$$

The collision matrix $S$ in moment space is the diagonal matrix

$$S = diag(s_1, s_2, s_3, s_4, s_5, s_6, s_7, s_8, s_9, s_{10}, s_{11}, s_{12}, s_{13}, s_{14}, s_{15}, s_{16}, s_{17}, s_{18}, s_{19}) \quad (23)$$

With $s_9$ equaling to $s_{13}$, the equilibrium moments are:

$$\mathbf{m}^{eq} = \left(\rho, -11\rho + 19\frac{j_x^2 + j_y^2 + j_z^2}{\rho_0}, 3\rho - \frac{11}{2}\frac{j_x^2 + j_y^2 + j_z^2}{\rho_0}, j_x, -\frac{2}{3}j_x, j_y, -\frac{2}{3}j_y, j_z, -\frac{2}{3}j_z, \right.$$
$$\left. \frac{2j_x^2 - (j_y^2 + j_z^2)}{\rho_0}, -\frac{1}{2}\frac{2j_x^2 - (j_y^2 + j_z^2)}{\rho_0}, \frac{j_y^2 - j_z^2}{\rho_0}, -\frac{1}{2}\frac{j_y^2 - j_z^2}{\rho_0}, \frac{j_x j_y}{\rho_0}, \frac{j_y j_z}{\rho_0}, \frac{j_x j_z}{\rho_0}, 0, 0, 0\right)^T \quad (24)$$

where

$$j_x = \rho u_x, \ j_y = \rho u_y, \ j_z = \rho u_z \quad (25)$$

The constant $\rho_0$ in Eq. (24) is the mean density in the system and it is usually set to be unity.

Taking $\rho_0$ in to account can reduce compressible effect in the model [39]. Then the unknown parameters in collision matrix $S$ are:

$$\begin{cases} s_1 = s_4 = s_6 = s_8 = 1.0, \ s_2 = 1.19, \ s_3 = s_{11} = s_{13} = 1.4, \ s_5 = s_7 = s_9 = 1.2 \\ s_{17} = s_{18} = s_{19} = 1.98, \ s_{10} = s_{12} = s_{14} = s_{15} = s_{16} = 1/(3\nu + 0.5) \end{cases} \quad (26)$$

The velocity field is solved using this D3Q19-MRT model. Non-slip boundary conditions in this article are fulfilled using bounce-back boundary conditions [10].

## 3.2 D3Q7-MRT-LBM Model for Heat Transfer

Yoshida and Nagaoka [40] proposed a Multiple-relaxation-time lattice Boltzmann model for convection and anisotropic diffusion equation. In this D3Q7 model, seven discrete velocities are needed for a three-dimensional problem. Li et al. [41] discussed boundary conditions for this thermal LBM model. This model has not been involved by any DDF model in LBM. In this article, we propose a double MRT model for fluid flow and heat transfer problem simulation. Velocity and temperature fields are solved with D3Q19-MRT and D3Q7-MRT, respectively.

D3Q7-MRT model is valid to solve the following standard convection-diffusion equation.

$$\frac{\partial \phi}{\partial t} + \frac{\partial}{\partial x_j}(v_j \phi) = \frac{\partial}{\partial x_i}\left(D_{ij} \frac{\partial \phi}{\partial x_j}\right) \quad (27)$$

where $\phi$ is a scalar variable and $D_{ij}$ is the diffusion coefficient. Energy equation for the benchmark problem in this article is shown in Eq. (5), which is a special case of Eq. (27). Therefore, D3Q7 model can be used to solve the energy equation in this article.

Each computing nodes have seven discrete velocities shown in Fig. 3:

$$\mathbf{u}_i = c \begin{bmatrix} 0 & 1 & -1 & 0 & 0 & 0 & 0 \\ 0 & 0 & 0 & 1 & -1 & 0 & 0 \\ 0 & 0 & 0 & 0 & 0 & 1 & -1 \end{bmatrix} \quad (28)$$

Similar to the density distribution, energy distribution $g_i$ can be obtained by:

$$g_i(\mathbf{r} + \mathbf{u}_i \Delta t, t + \Delta t) - g_i(\mathbf{r}, t) = -N^{-1} \cdot Q \cdot [n_i(\mathbf{r}, t) - n_i^{eq}(\mathbf{r}, t)], \quad i = 1, 2, \ldots 7 \quad (29)$$

where $n_i^{eq}(\mathbf{r}, t)$ is the equilibrium function for $n_i(\mathbf{r}, t)$, $N$ and $Q$ are the transform matrix and collision matrix for the energy distribution [40].

Macroscopic parameters relate to the energy distributions as:

$$\begin{cases} T = \sum_{i=1}^{7} g_i(\mathbf{r}, t), \\ \mathbf{n} = N \cdot [g_1, \ g_2, \ \cdots \ g_7]^T \end{cases} \quad (30)$$

The energy weight factors $\omega_{T,i}$ are:

$$\omega_{T,i} = \begin{cases} 1/4, & (i=1), \\ 1/8, & (i=2,3,\cdots,7) \end{cases} \tag{31}$$

Transform matrix in D3Q7 model is defined as:

$$N = \begin{bmatrix} 1 & 1 & 1 & 1 & 1 & 1 & 1 \\ 0 & 1 & -1 & 0 & 0 & 0 & 0 \\ 0 & 0 & 0 & 1 & -1 & 0 & 0 \\ 0 & 0 & 0 & 0 & 0 & 1 & -1 \\ 6 & -1 & -1 & -1 & -1 & -1 & -1 \\ 0 & 2 & 2 & -1 & -1 & -1 & -1 \\ 0 & 0 & 0 & 1 & 1 & -1 & -1 \end{bmatrix} \tag{32}$$

and its corresponding equilibrium moments are:

$$\boldsymbol{n}^{eq} = \begin{bmatrix} T, & uT, & vT, & wT, & aT, & 0, & 0 \end{bmatrix}^T \tag{33}$$

where $a$ is a constant relating to the coefficient $\omega_{T,1}$ by:

$$a = (7\omega_{T,1} - 1) = 3/4 \tag{34}$$

The definition of collision matrix $\boldsymbol{Q}$ is:

$$\boldsymbol{Q}^{-1} = \begin{bmatrix} \tau_1 & 0 & 0 & 0 & 0 & 0 & 0 \\ 0 & \bar{\tau}_{xx} & \bar{\tau}_{xy} & \bar{\tau}_{xz} & 0 & 0 & 0 \\ 0 & \bar{\tau}_{xy} & \bar{\tau}_{yy} & \bar{\tau}_{yz} & 0 & 0 & 0 \\ 0 & \bar{\tau}_{xz} & \bar{\tau}_{yz} & \bar{\tau}_{zz} & 0 & 0 & 0 \\ 0 & 0 & 0 & 0 & \tau_5 & 0 & 0 \\ 0 & 0 & 0 & 0 & 0 & \tau_6 & 0 \\ 0 & 0 & 0 & 0 & 0 & 0 & \tau_7 \end{bmatrix} \tag{35}$$

The off-diagonal components correspond to the rotation of principal axis of anisotropic diffusion [40]. The relaxation coefficients $\bar{\tau}_{ij}$ are related to the diffusion coefficient matrix by:

$$\bar{\tau}_{ij} = \frac{1}{2}\delta_{ij} + \frac{\Delta t}{\varepsilon \Delta x^2} D_{ij} \tag{36}$$

where $\varepsilon$ is a constant 0.25 in a three-dimensional problem, and $\delta_{ij}$ is the Kronecker's delta:

$$\delta_{ij} = \begin{cases} 0 & \text{if } i \neq j, \\ 1 & \text{if } i = j. \end{cases} \tag{37}$$

The relaxation coefficient $\tau_1$ for the conserved quantity does not affect the numerical solution, and $\tau_5$, $\tau_6$ and $\tau_7$ only affect error terms. After testing, this article use 1 for these four coefficients.

Temperature field is solved using this D3Q7-MRT model and thermal boundary conditions are solved based on the settings in Ref. [41].

## 4. Results and discussions

Lattice velocity $c$ is always set as unity in LBM. Therefore, parameters in any lattice unit are also non-dimensional. For a natural convection problem, Ref. [28] describes detailed settings for parameters in lattice unit.

$$\begin{cases} X = \dfrac{x}{H}, Y = \dfrac{y}{H}, Z = \dfrac{z}{H}, u_c = \sqrt{g\beta(T_h - T_c)H}, Ma = \dfrac{u_c}{c_s}, U = \dfrac{u}{\sqrt{3}c_s}, V = \dfrac{v}{\sqrt{3}c_s}, \\ W = \dfrac{w}{H}, \tau = \dfrac{t \cdot \sqrt{3}c_s}{H}, \theta = \dfrac{T - T_c}{T_h - T_c}, P = \dfrac{p}{3\rho c_s^2}, Pr = \dfrac{\nu}{\alpha}, Ra = \dfrac{g\beta(T_h - T_c)H^3 Pr}{\nu^2} \end{cases} \quad (38)$$

Mach number $Ma$, Pandtl number $Pr$ and Rayleigh number $Ra$ are the character parameters.

Natural convection problem is fully defined with $Pr$ and $Ra$. LBM includes the speed of sound, $c_s$. So we have to include $Ma$ to fulfill this non-dimensional process for lattice unit.

Wang et al. [42] demonstrated that $Ma$ has little effect on accuracy of MRT simulation. Incompressible air is the working fluid. $Ma$ is 0.1 while $Pr$ is 0.71 for all the cases in this article. Three types of natural convections in Section 2 are solved for various Rayleigh numbers ranged from $1\times10^4$ to $1\times10^5$.

The two vertical walls ($Y=0$ and $Y=1$) are kept at $T_h$ and $T_c$ for all the cases in consideration. The local Nusselt number is defined as

$$Nu = \dfrac{h}{k/H}\bigg|_{Y=0} = -\dfrac{\partial \theta}{\partial Y}\bigg|_{Y=0} \quad (39)$$

The average Nusselt number $\overline{Nu_{3D}}$:

$$\overline{Nu_{3D}} = \int_0^1 \int_0^1 (Nu)\,dX\,dZ \quad (40)$$

and the maximum Nusselt number $Nu_{max}$ on the heat wall are important parameters to discuss three-dimensional natural convection problems. Due to the symmetry of the cubic natural convection, the mid-plane of the cubic (X=0.5) plays an important role in this problem and the average Nusselt number $\overline{Nu_{2D}}$ at mid-plane is also in consideration.

$$\overline{Nu_{2D}} = \int_0^1 (Nu_{X=0.5}) dZ \tag{41}$$

Besides $\overline{Nu_{2D}}$, $\overline{Nu_{3D}}$ and $Nu_{max}$, the maximum velocities in all directions are also discussed.

## 4.1 Type 1 natural convection

For type 1 natural convection, two different sets of grids ($50\times50\times50$ and $60\times60\times60$) are employed. Refs. [31, 32, 33 and 43] reported $\overline{Nu_{3D}}$ for $Ra = 1\times10^4$ and $Ra = 1\times10^5$. Table 1 shows the comparison between present results and that in references. These references results agree with each other well and their averages (2.07 for $Ra = 1\times10^4$ and 4.36 for $Ra = 1\times10^5$) can be viewed as standard results. For the case of $Ra = 1\times10^4$, the present numerical results in different grids are both close to the standard one. For the case of $Ra = 1\times10^5$, however, the results obtained using grid of $60\times60\times60$ agreed better with the results in the literature. Refs. [32 and 35] reported $\overline{Nu_{2D}}$ for this type of natural convection. Table 2 is the comparison between results obtained from the present LBM and those from the references. The mean values from the references (2.28 for $Ra = 1\times10^4$ and 4.64 for $Ra = 1\times10^5$) are taken as standard ones. For the case of $Ra = 1\times10^4$, the results from the two grid number are the same and their differences between standard one are negligible. For the case of $Ra = 1\times10^5$, the differences between two the results from the two grid numbers and standard one are within 2% and the result from grid number of $60\times60\times60$ grids is closer to the standard one. Refs. [31 and 34] reported the $Nu_{max}$ and maximum velocities, respectively. Non-dimensional process in these references is different from that in article. We can get the velocity $U_s$ in reference unit with the numerical results $U_l$ using the following equation:

$$U_s = U_l \sqrt{3\Pr\cdot Ra} / Ma \tag{42}$$

Tables 3 to 5 are the comparisons between numerical and reference results. They indicate that the results from different grid numbers agreed well with those from the references. The above comparisons indicated that $\overline{Nu_{2D}}$, $\overline{Nu_{3D}}$, $Nu_{max}$ and maximum velocities in different directions results all agreed well with reference ones; thus the proposed double MRT model is valid for the Type 1 cubic natural convection simulation. Considering the computational efficiency and accuracy, the grid number of $50\times50\times50$ is suitable for the case of $Ra=1\times10^4$ while $60\times60\times60$ more appropriate for thee case of $Ra=1\times10^5$. The other two types of natural convection simulations for various Rayleigh numbers also have the same grid settings. Temperature and velocity fields for the cubic natural convection are very important. But few references include three-dimensional visual results. Figure 4 shows the temperature field for Type 1 problem at $Ra=1\times10^4$. Surface temperature distribution and temperature isosurfaces are included in Figs. 4 (a) and (b). The temperature turns to be higher with increasing z in the cubic cavity. Convection has dominated the heat transfer process and temperature isosurfaces does not change a lot in the x-direction. Figure 4 (c) to (e) show the temperature distributions on different locations for Type 1 problem. Regarding the boundary settings, it is common to argue the working condition on the mid-plane of the cubic (X=0.5) can be viewed as a two-dimensional problem [31]. Temperature field (X=0.5) agrees well with the two-dimensional results in Ref. [28]. In the other two locations (Y=0.5 and Z=0.5), isothermal lines are almost parallel to the X-axis. It supports the two-dimensional assumption. Figure 4 (f) shows the Nusselt number distribution on the hot surface (Y=0). Nusselt numbers at the mid-plane of the cubic (X=0.5) are higher than that in the other regions due to the side wall effect to these three-dimensional problem. Non-slip condition is applied to all the boundaries and it slows down the convection flow near the boundaries. Consequently, the convection effect to heat transfer is also lowered.

The velocity field in the cubic cavity is also discussed. Figure 5 shows the streamtraces result for $Ra=1\times10^4$. Its main tendency is a two-dimensional flow in the YZ plane. It also has a tendency to flow to the center of the cavity in the X-direction. Figure 5 (b)-(d) show the streamtraces on different locations. The results on the mid-plane of the cubic (X=0.5) agreed well with two-dimensional ones in Ref. [28]. The results on Y=0.5 show that the fluid have a tendency to flow to the cavity center in the X-direction. Meanwhile, two intersections exist for the streamtraces on the surface Z=0.5. It indicates fluid flowing to the cavity center in the X-direction joins the two-dimensional flow in the Y and Z directions on the mid-plane of the cubic (X=0.5).

Type 1 problem is also discussed for $Ra=1\times10^5$. Figure 6 shows its temperature results. Convection effect is stronger comparing with the case at lower Rayleigh number. And temperature difference between top and bottom of the cavity turns to be greater. The temperature isosurfaces' changes in the X-direction are limited. Figure 6 (c)-(e) are temperature distributions on different locations. Because of the boundary settings and symmetry of this problem, the working condition on mid-plane of the cubic (X=0.5) is still close to a two-dimensional one. The temperature field for X=0.5 agreed well with the two dimensional result in Ref. [28]. Temperature distributions for Y=0.5 and Z=0.5 also prove that the two-dimensional assumption on the mid-plane of the cubic (X=0.5) is reasonable. Nusselt number distribution on the hot surface is shown in Fig. 6 (f). It decreases with increasing Z. Its isolines are almost parallel to the X axis except the bottom region. In that region, Nusselt number turns to be higher when closing to the mid-plane of the cubic (X=0.5). As shown in Fig. 7 (a), the velocity field is more complicated and flow is stronger than that of the case of $Ra = 1 \times 10^4$.

The two-dimensional flow in the Y- and Z- directions is still the main tendency. Figure 7 (b)-(d) include the surface streamtraces on different locations. Results on the mid-plane of the cubic (X=0.5) agree well with the two-dimensional ones in Ref. [28]. On the surface of Y=0.5, fluid flows to the center in the X-direction. Four streamtrace intersections exist on the surface (Z=0.5). It indicates a stronger three-dimensional effect to the fluid flow.

Type 1 cubic natural convection is widely used as a benchmark problem to test numerical methods for three-dimensional fluid flow and heat transfer simulations. The above results show that the proposed double MRT model is reliable for this kind of problem. Meanwhile, few references include three-dimensional visual results. Since type 1 problem is not physically-realizable [30], we will continue to discuss type 2 cubic natural convection.

## 4.2 Type 2 natural convection

It is physically-realizable regarding its boundary condition settings in Section 2. Leong et al. [30] obtained the experimental results of $\overline{Nu_{3D}}$ for this type of natural convection. For $Ra = 1 \times 10^4$ and $Ra = 1 \times 10^5$, the present $\overline{Nu_{3D}}$ results agree with Ref. [30] ones well shown in Table 6. It also proves that the proposed double MRT model is valid for three-dimensional fluid flow and heat transfer simulation. More detailed results are included for type 2 cubic natural convection as benchmark solutions.

Figure 8 shows the temperature results for Type 2 problem at $Ra = 1 \times 10^4$. Hot and cold surfaces (Y=0 and Y=1) are kept at constant temperatures. The remained four side walls have linear temperature distributions in the Y-direction. The temperature isosurfaces show that convection dominates the heat transfer process. Temperature differences in the X-direction are significant. It indicates that type 2 problem has clear 3-D characteristics. Temperature distributions on different locations are shown in Figs. 8 (c)-(e). Results on the mid-plane of the cubic (X=0.5) show the convection effect. Results on the surfaces (Y=0.5 and Z=0.5) indicate that temperature differences at the center of the cavity are not significant. Figure 8 (f) shows the Nusselt number distribution on the hot surface for this working condition. The location at which maximum Nusselt number is reached is higher in the cavity than that in Type 1 problem with the same Rayleigh number. Nusselt number at the mid-plane of the cubic (X=0.5) can be lower than that in the other locations at the same height. From Fig. 9 (a), we can find two-dimensional flow in the Y- and Z-directions. And the flow in the X-direction is also strong. Streamtraces on different locations are shown in Figs. 9 (b)-(d). One vortex locates on the mid-plane of the cubic (X=0.5). It is quite similar to that in Type 1 problem for $Ra = 1 \times 10^4$. The surface (Y=0.5) has four symmetry vortexes. Streamtraces on the surface (Z=0.5) have two intersections. They show the flow tendency in all directions.

Convection effect is more valid in Type 2 cubic natural convection when Rayleigh number is $1\times10^5$ as shown in Fig. 10. Nusselt number isolines in Fig. 10 (f) have similar tendency as that in Fig. 8 (f). The difference is that Nusselt numbers are higher due to the stronger convection effects. All three cases discussed above (Type 1 problem for $Ra=1\times10^4$ and $Ra=1\times10^5$; Type 2 problem for $Ra=1\times10^4$) all have strong two-dimensional flows in the Y- and Z-directions. Cubic cavity streamtraces in Fig. 11 shows this two-dimensional flow is not as strong as that in the other cases. Moreover, Figure 11 (b)-(d) include streamtraces on different surfaces. Mid-plane of the cubic (X=0.5) have two vortexes while two and four streamtraces intersections exists on the surfaces (Y=0.5 and Z=0.5), respectively. Character factors for type 2 problem with different Rayleigh numbers are shown in Table 7, which can be used as benchmark solutions.

Comparing with Type 1 problem, Type 2 cubic natural convection has three advantages to be a benchmark problem to test numerical method for a three-dimensional fluid flow and heat transfer: (1) it is physically-realizable, (2) it has experimental results that agree well with the present numerical ones, and (3) three-dimensional effect is more valid in type 2 problem. On the other hand, type 2 temperature isosurfaces do not change a lot in the X-direction at the region close to the cubic cavity top as can be seen Figs. 8 and 10. To discuss a real three-dimensional problem, we propose Type 3 cubic natural convections.

## 4.3 Type 3 natural convection

In type 3 cubic natural convection, surfaces (Y=0 and Y=1) have constant temperatures, and side walls (X=0 and X=1) have linear temperature distributions in the Y–direction while the top and bottom of the cubic are kept adiabatic. We discuss the type 3 problem for $Ra=1\times10^4$ first. Figure 12 (a)-(e) show the temperature field and temperature distribution on different locations for this case and three-dimensional features are clearly shown. Nusselt number distribution in Fig. 12 (f) is similar to that in Type 1 problem shown in Fig. 4 (f). For the same Rayleigh number, Nusselt numbers are lower than that in Type 1 problem and higher than that in Type 2 problem. Boundary with linear temperature distribution lowers the convection effect, comparing with the adiabatic condition. Streamtraces shown in Fig. 13 indicate the flow in the X-, Y- and Z-directions are all strong. Figure 13 (b)-(d) include the streamtraces on different locations. Mid-plane of the cubic (X=0.5) has one vortex, four vortexes exist on the surface (Y=0.5), and streamtraces on the surface (Z=0.5) have two intersections. These results are similar to that in Type 2 problem for $Ra=1\times10^4$.

Type 3 problem is then discussed for $Ra=1\times10^5$ and the results are shown in Figs.14 and 15. They show more complicated velocity and temperature fields, which have clear three-dimensional features. Temperature isosurfaces changes significantly in the X-, Y- and Z-directions. For the same Rayleigh number, Nusselt numbers for Type 3 problem are still lower than that in Type 1 problem and higher than that in Type 2 problem. Streamtraces results indicate flow in all directions are strong. It shows clear 3-D characteristics in type 3 problem. Table 8 records the characteristic quantities for various Rayleigh numbers. They can be used as benchmark solutions for Type 3 problem.

# 5. Conclusions

Three-dimensional double MRT model is proposed for LBM for fluid flow and heat simulation. Three types of cubic natural convection problems with various Rayleigh numbers are solved with the proposed method. Temperature field, hot surface Nusselt number distribution, velocity field, $\overline{Nu_{2D}}$, $\overline{Nu_{3D}}$, $Nu_{max}$ and maximum velocities in different directions are discussed.

The results of Type 1 problem agreed well with the reference ones, and the results from Type 2 problem fit the reported experimental results well. Therefore, the proposed double MRT is valid for three-dimensional simulation. Type 2 problems are more physically-realizable comparing with the type 1 problems. Their numerical results are reported for the first time. Type 3 problems are also investigated because their results have more general three-dimensional features. All these three types' 3-D natural convection results can be used as benchmark solutions for further researches.

**Acknowledgement**
Support for this work by Chinese National Natural Science Foundations under Grants 51129602 and 51476103, and Innovation Program of Shanghai Municipal Education Commission under Grant 14ZZ134 are gratefully acknowledged.

23, pp. 3817-3828, 1997.
[39] X. He and L. Luo, Lattice Boltzmann model for the incompressible Navier-Stokes Equation, Journal of Statistical Physics, Vol. 88, No. 3-4, pp. 927-944, 1997.
[40] H. Yoshida and M. Nagaoka, Multiple- relaxation-time lattice Boltzmann model for the convection and anisotropic diffusion equation, Journal of Computational Physics, Vol. 229, No. 20, pp. 7774-7795, 2010.
[41] L. Li, R. Mei and J. Klausner, Boundary conditions for thermal lattice Boltzmann equation method, Journal of Computational Physics, Vol. 237, pp. 366-395, 2013.
[42] J. Wang, D. Wang, P. Lallemand and L. Luo, Lattice Boltzmann simulations of thermal convective flows in two dimensions, Computers and Mathematics with Applications, Vol. 65. pp. 262-286, 2013.
[43] T. Fusegi, J. Hyun, K. Kuwahara and B. Farouk, A numerical study of three-dimensional natural convection in a differentially heated cubical enclosure, International Journal of Heat and Mass Transfer, Vol. 34, No. 6, pp. 1543-1557, 1991.

# Figure Captions

Fig. 1 Cubic natural convection
Fig. 2 D3Q19 model
Fig.3 D3Q7 model

Fig. 4 Type 1, temperature results, $Ra = 1\times10^4$

Fig. 5 Type 1, Streamtrace results, $Ra = 1\times10^4$

Fig. 6 Type 1, temperature results, $Ra = 1\times10^5$

Fig. 7 Type 1, Streamtrace results, $Ra = 1\times10^5$

Fig. 8 Type 2, temperature results, $Ra = 1\times10^4$

Fig. 9 Type 2, Streamtrace results, $Ra = 1\times10^4$

Fig. 10 Type 2, temperature results, $Ra = 1\times10^5$

Fig. 11 Type 2, Streamtrace results, $Ra = 1\times10^5$

Fig. 12 Type 3, temperature results, $Ra = 1\times10^4$

Fig. 13 Type 3, Streamtrace results, $Ra = 1\times10^4$

Fig. 14 Type 3, temperature results, $Ra = 1\times10^5$

Fig. 15 Type 3, Streamtrace results, $Ra = 1\times10^5$

Table List

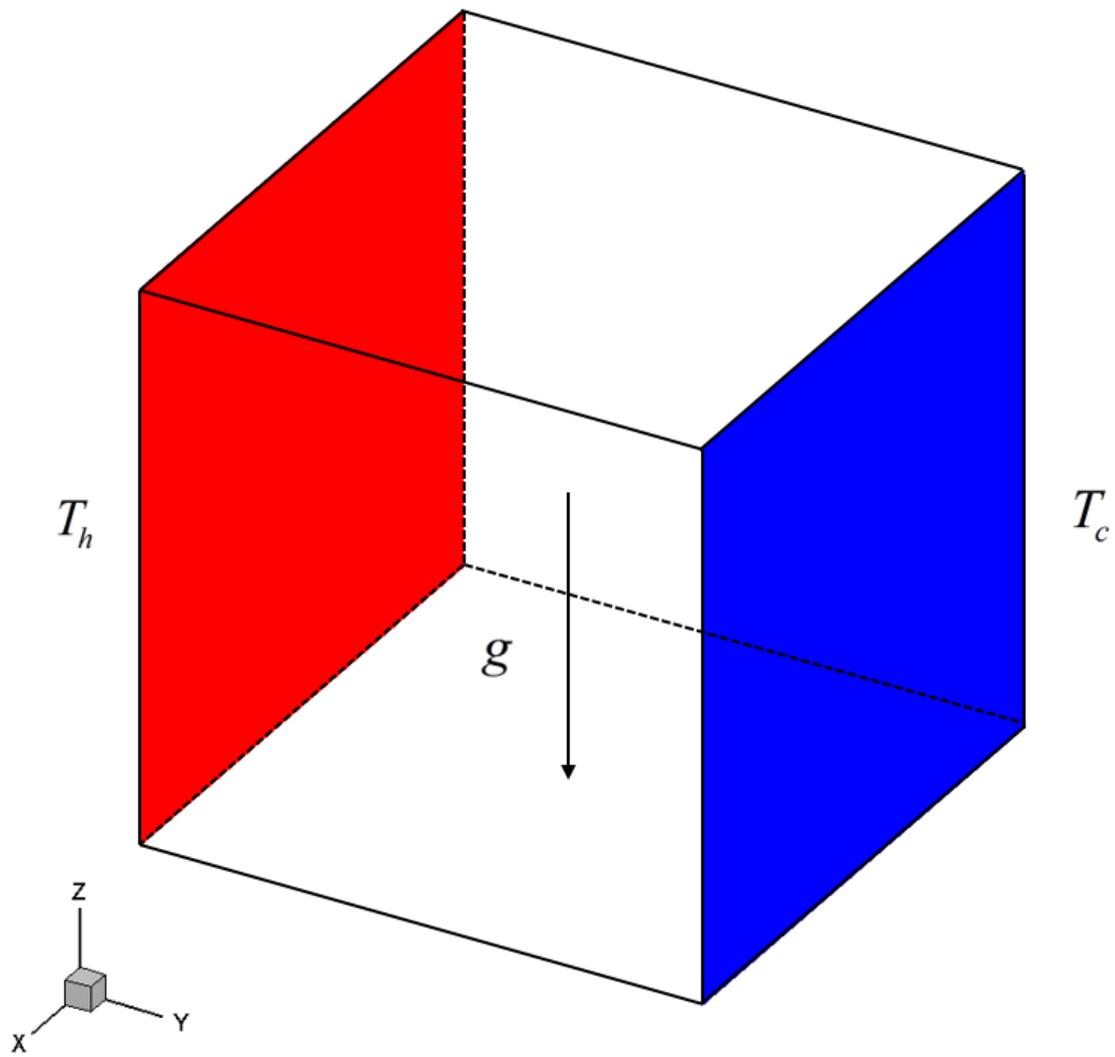

Fig. 1

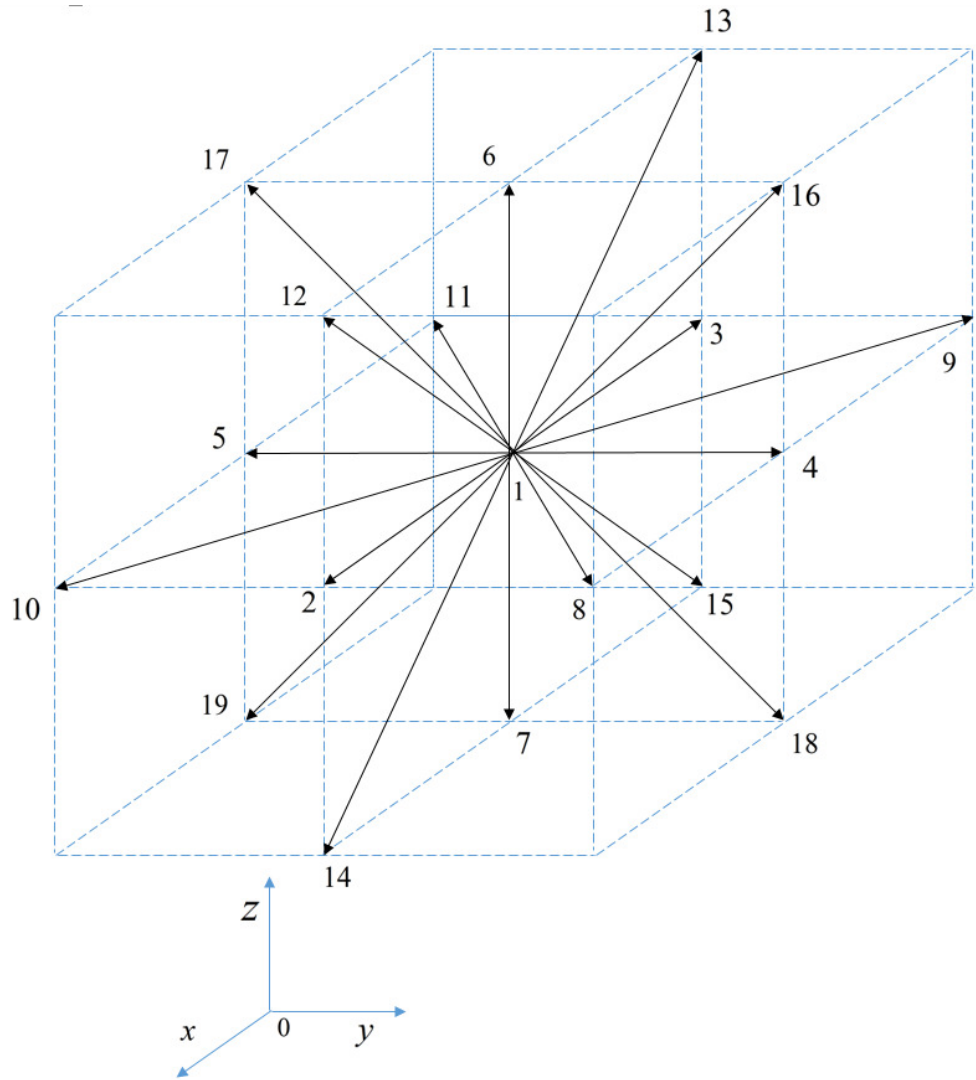

Fig. 2

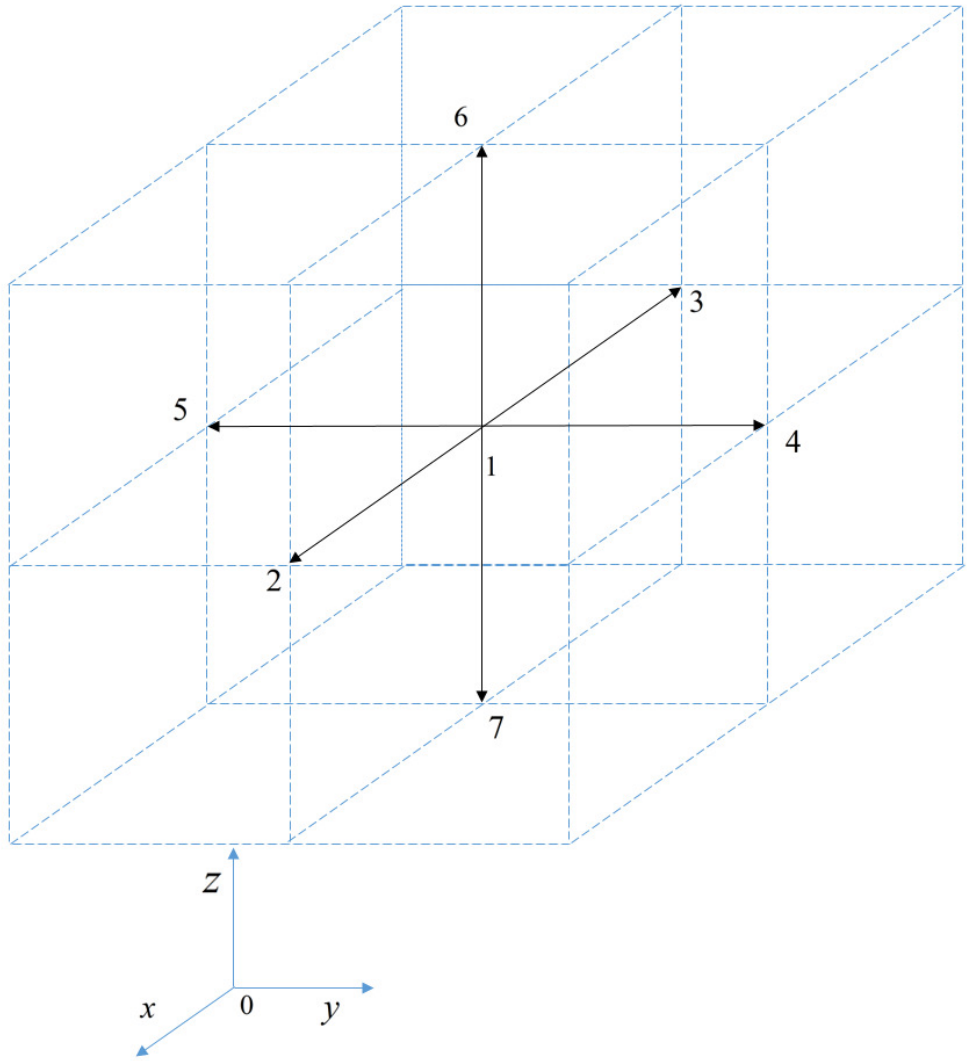

Fig. 3

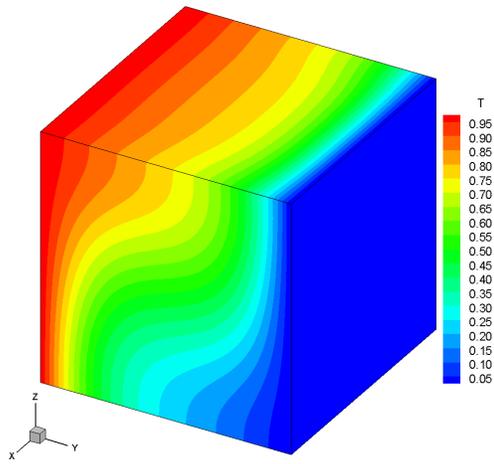
(a) Surface temperature distribution

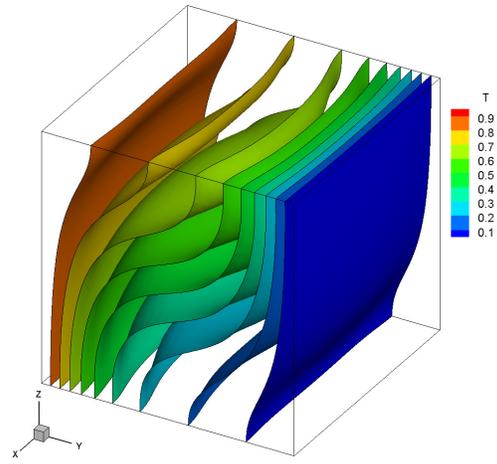
(b) Temperature isosurfaces

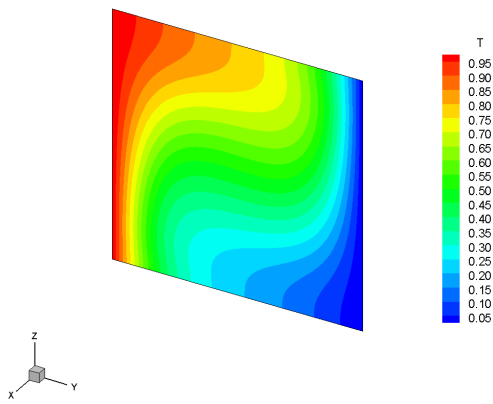
(c) X=0.5

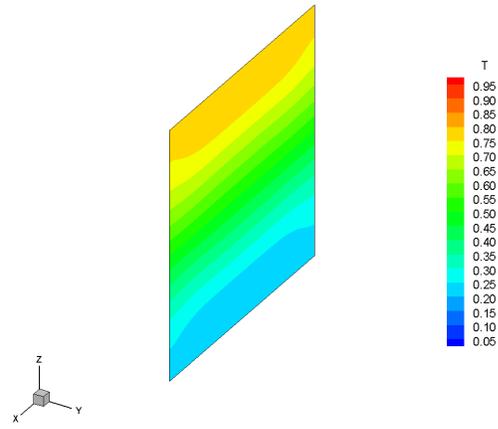
(d) Y=0.5

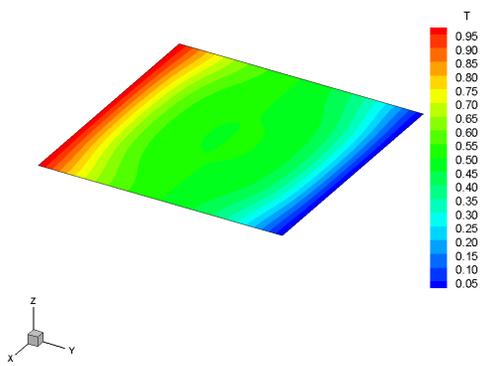
(e) Z=0.5

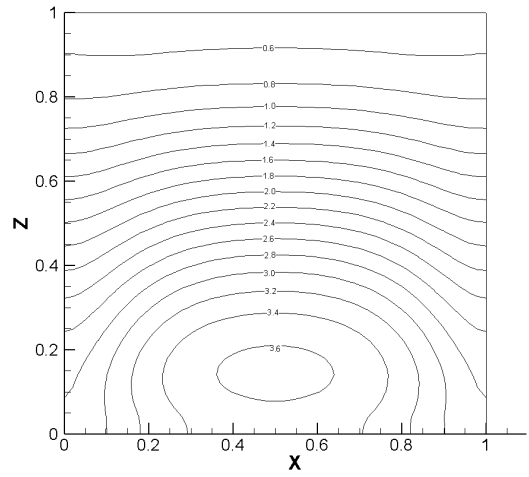
(f) Nusselt number distribution

Fig. 4

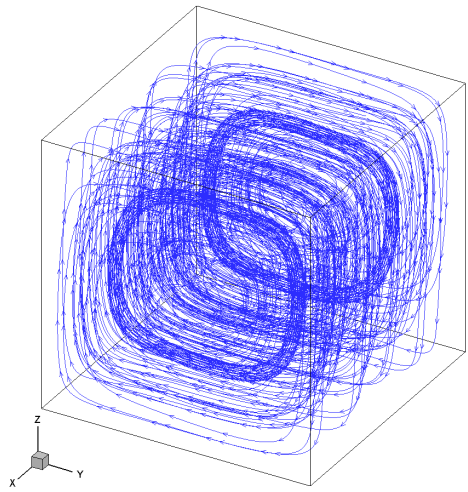
(a) 3D results

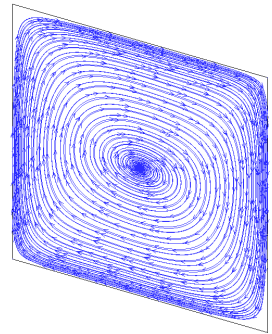
(b) X=0.5

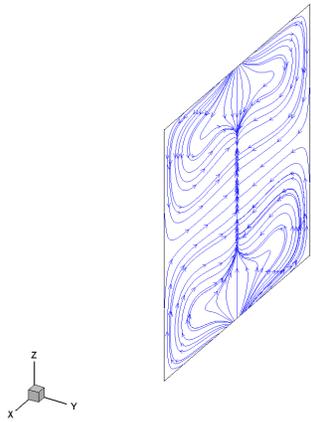
(c) Y=0.5

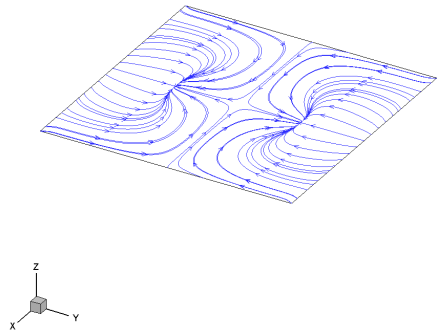
(d) Z=0.5

Fig. 5

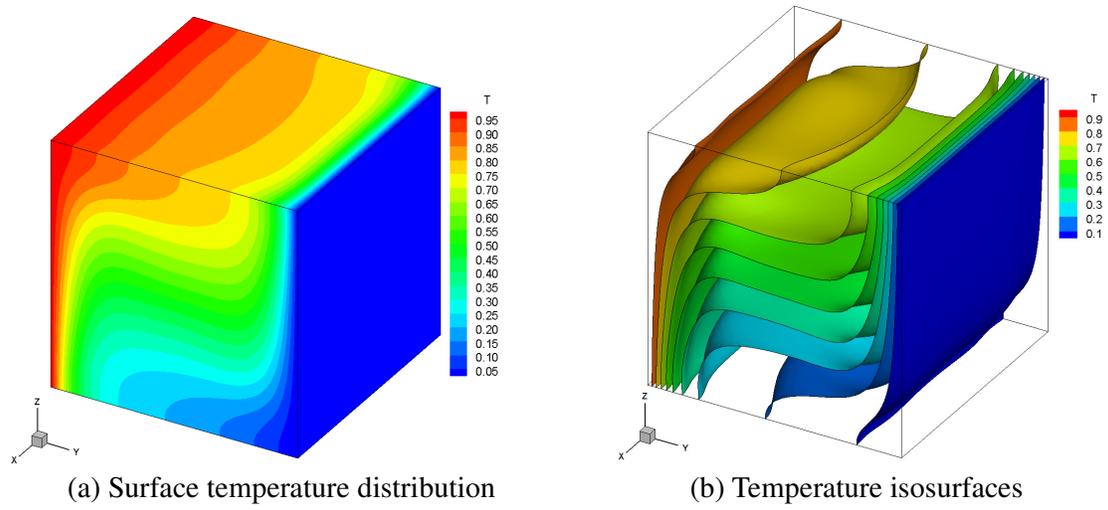

(a) Surface temperature distribution  (b) Temperature isosurfaces

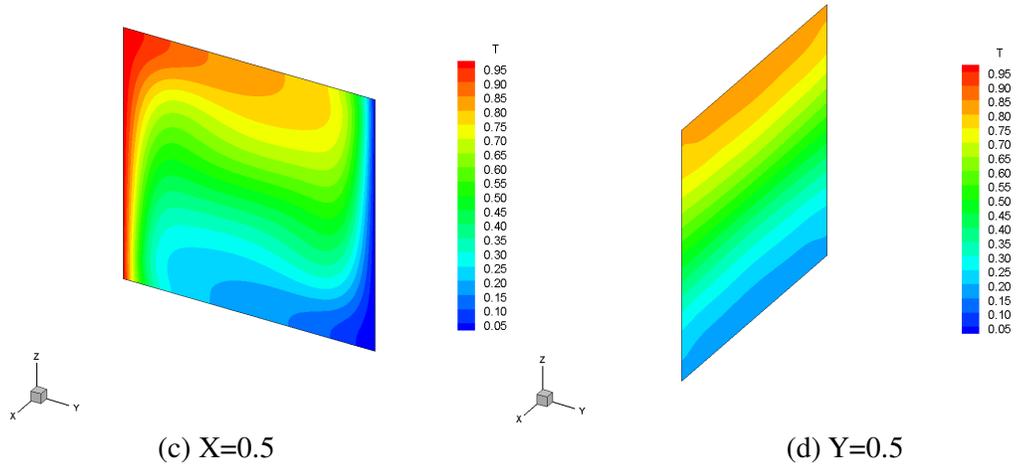

(c) X=0.5  (d) Y=0.5

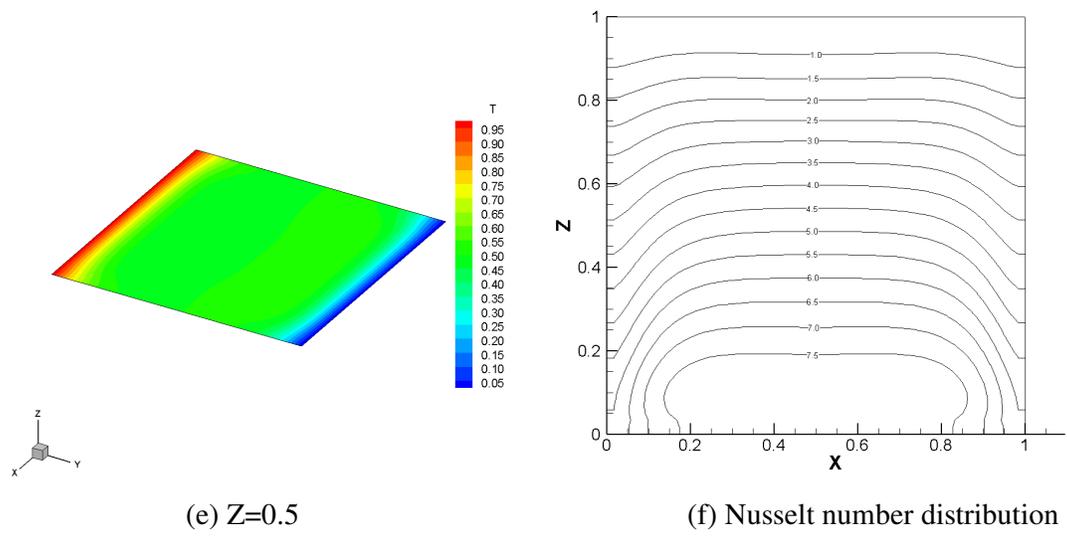

(e) Z=0.5  (f) Nusselt number distribution

Fig. 6

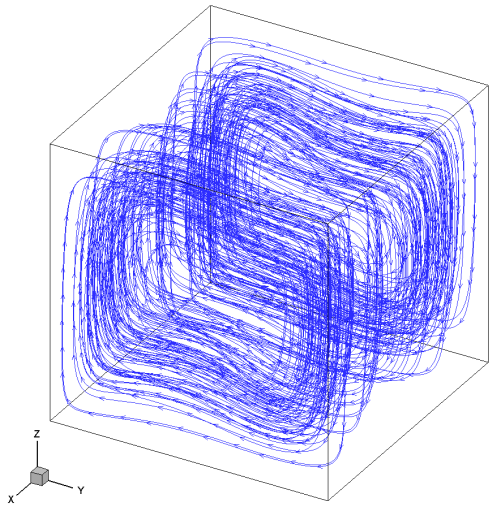
(a) 3D results

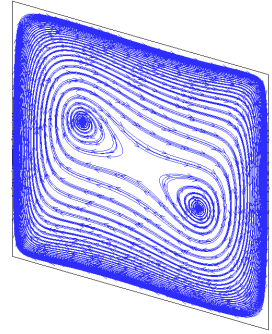
(b) X=0.5

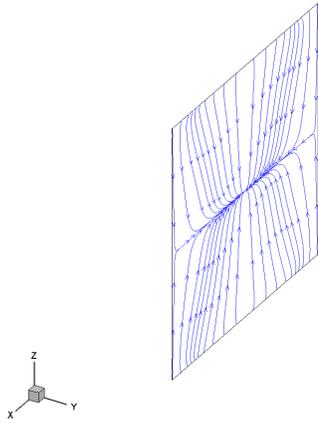
(c) Y=0.5

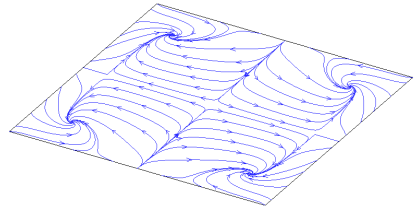
(d) Z=0.5

Fig. 7

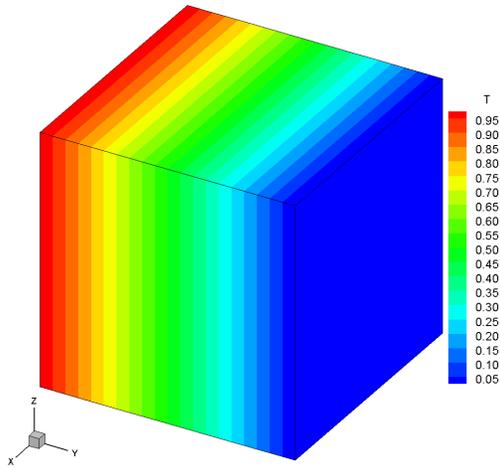

(a) Surface temperature distribution

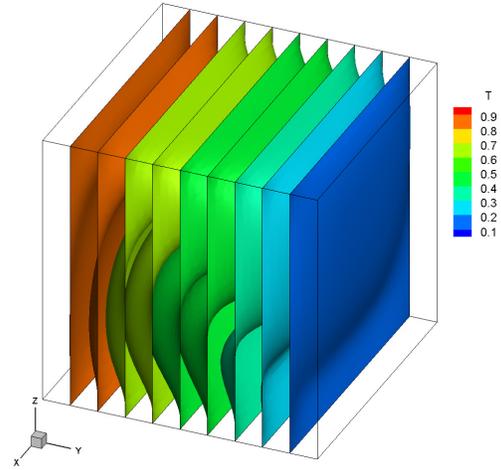

(b) Temperature isosurfaces

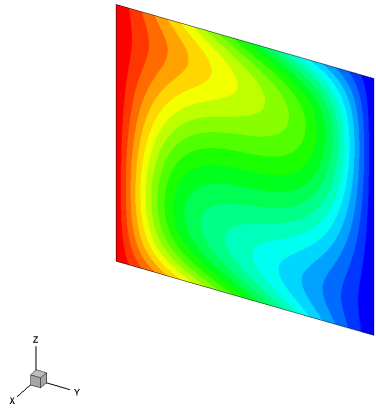

(c) X=0.5

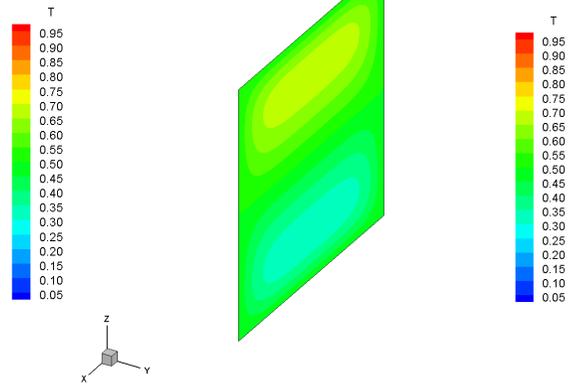

(d) Y=0.5

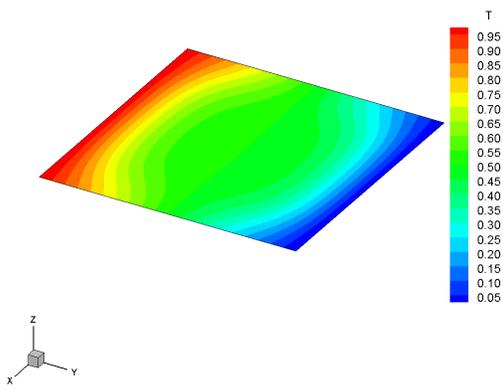

(e) Z=0.5

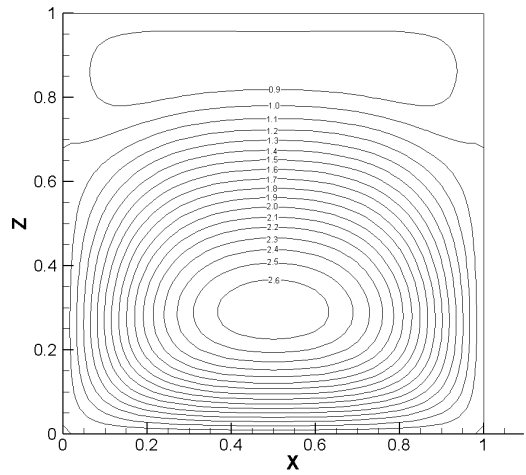

(f) Nusselt number distribution

Fig. 8

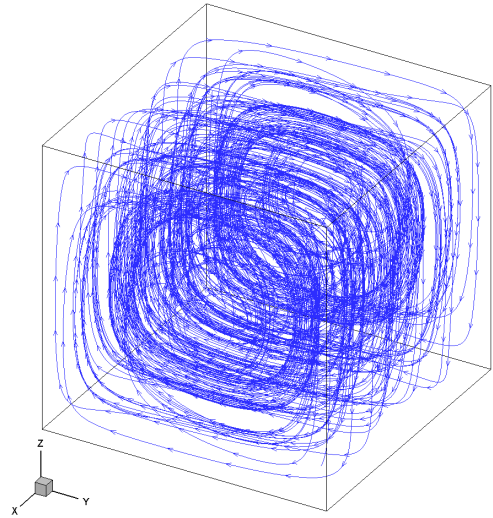

(a) 3D results

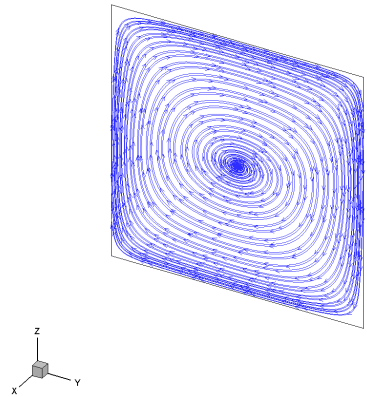

(b) X=0.5

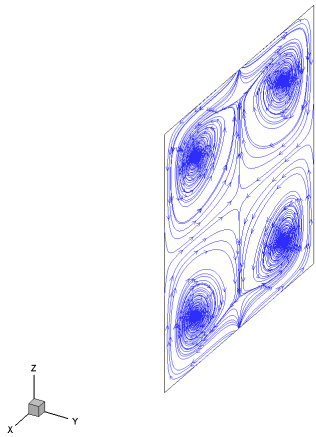

(c) Y=0.5

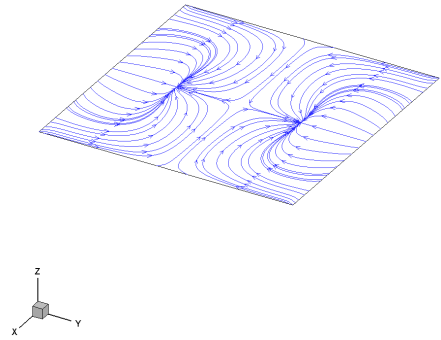

(d) Z=0.5

Fig. 9

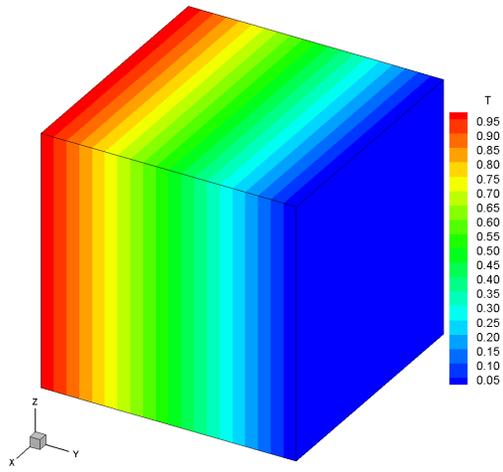
(a) Surface temperature distribution

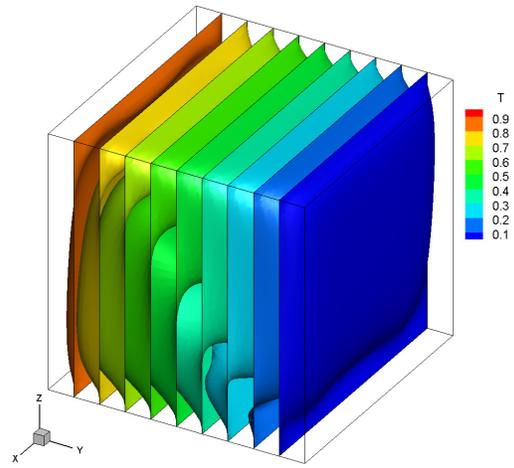
(b) Temperature isosurfaces

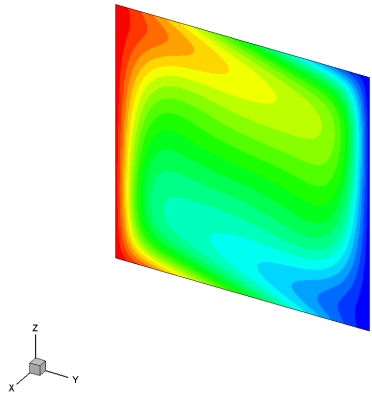
(c) X=0.5

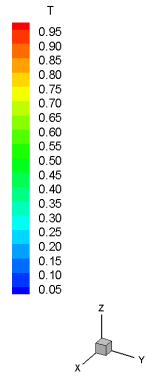
(d) Y=0.5

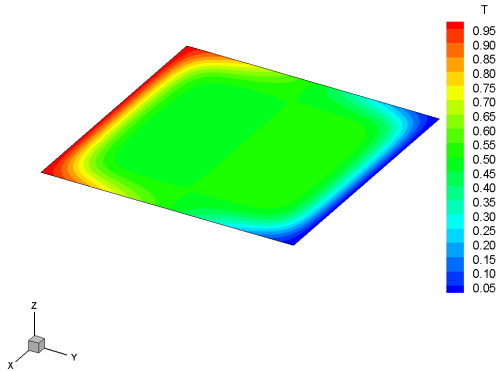
(e) Z=0.5

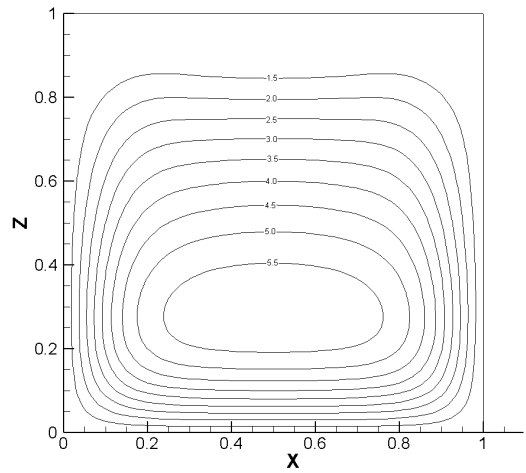
(f) Nusselt number distribution

Fig. 10

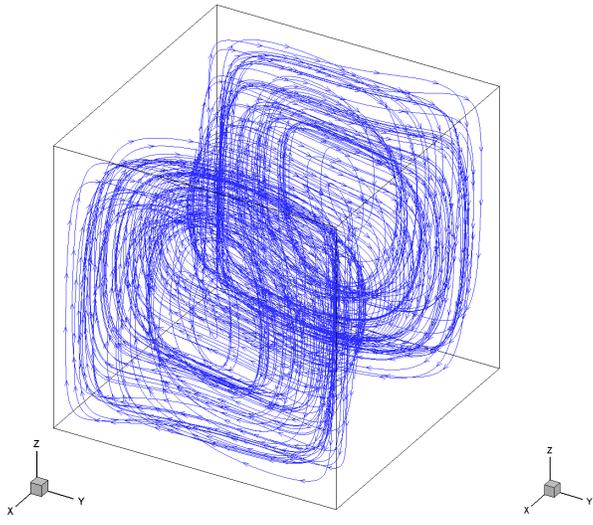

(a) 3D results

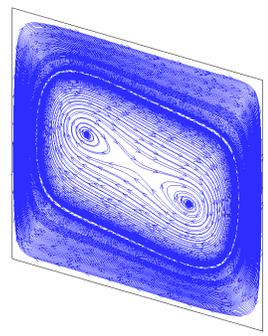

(b) X=0.5

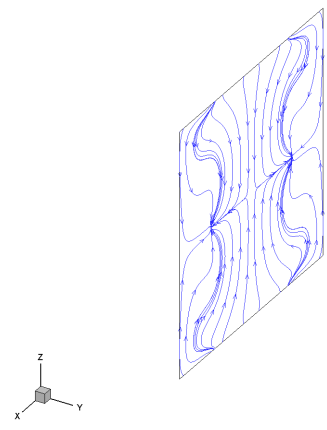

(c) Y=0.5

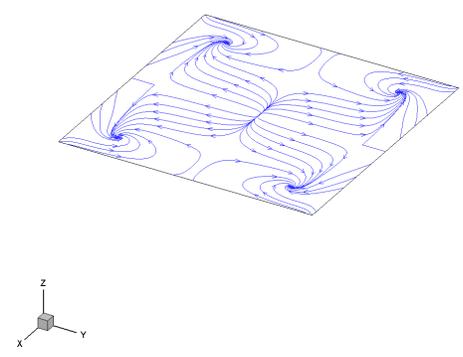

(d) Z=0.5

Fig. 11

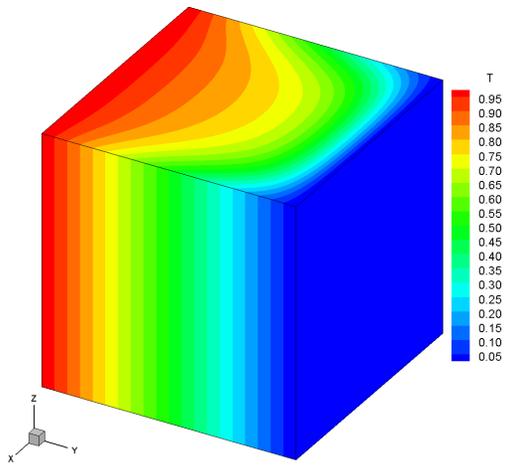
(a) Surface temperature distribution

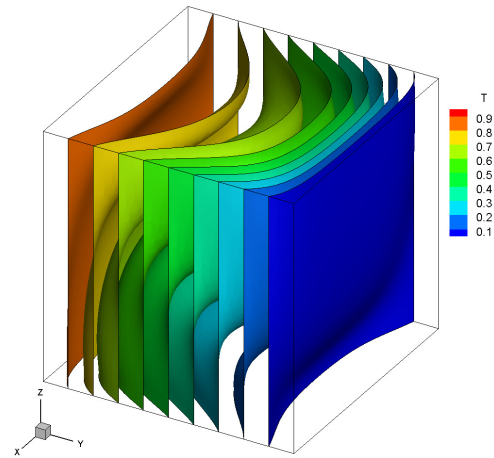
(b) Temperature isosurfaces

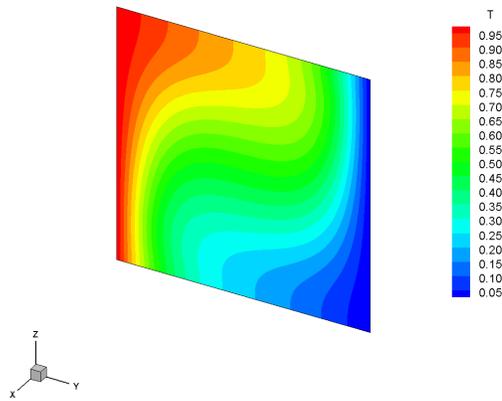
(c) X=0.5

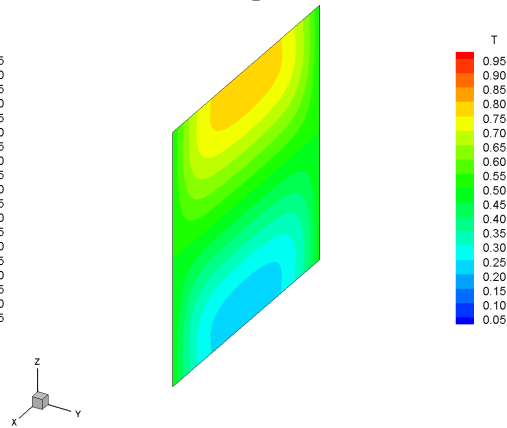
(d) Y=0.5

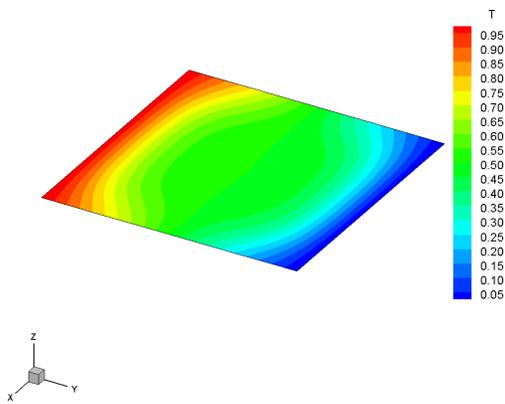
(e) Z=0.5

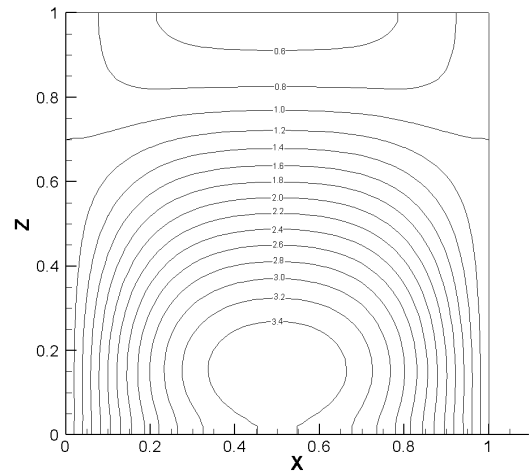
(f) Nusselt number distribution

Fig. 12

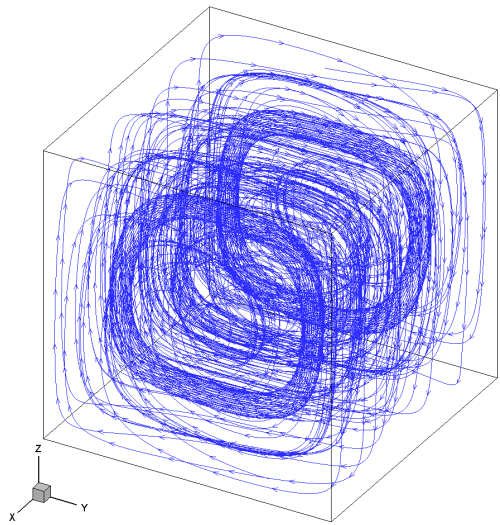
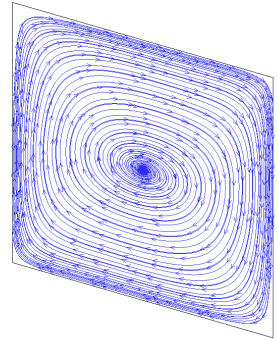

(a) 3D results  (b) X=0.5

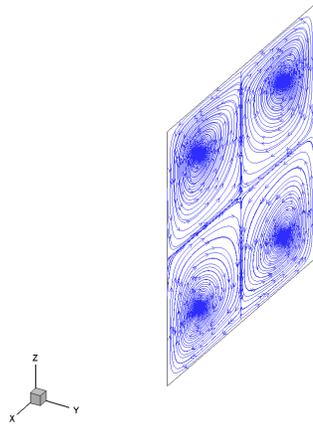
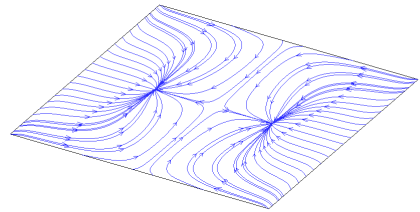

(c) Y=0.5  (d) Z=0.5

Fig. 13

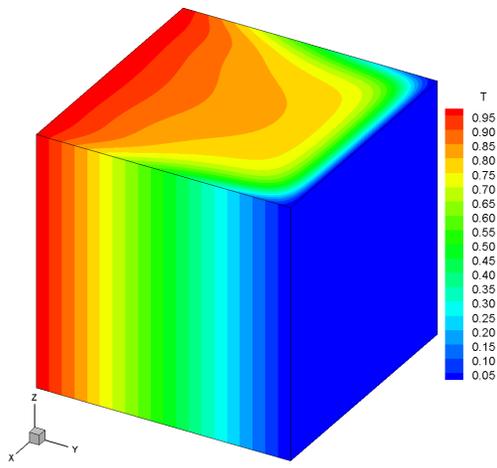
(a) Surface temperature distribution

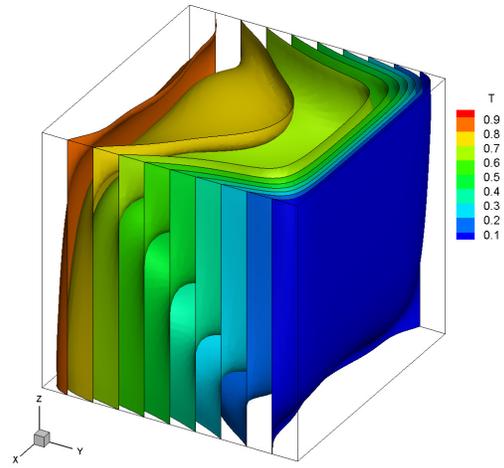
(b) Temperature isosurfaces

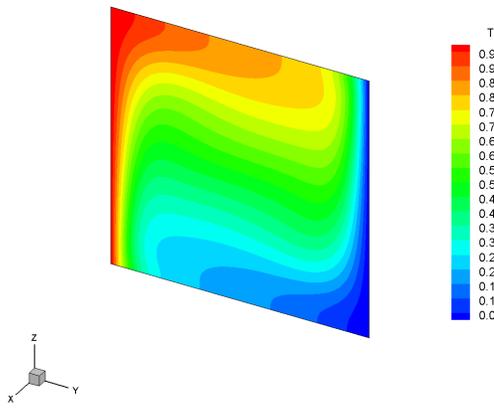
(c) X=0.5

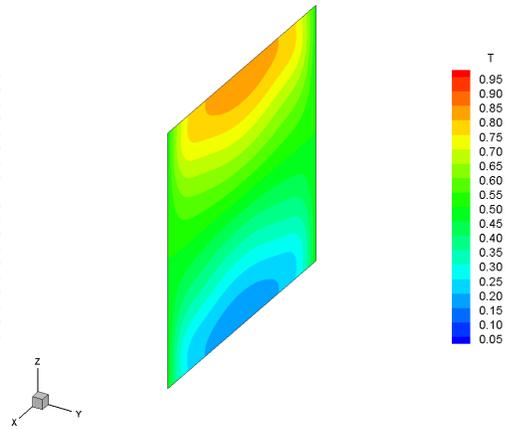
(d) Y=0.5

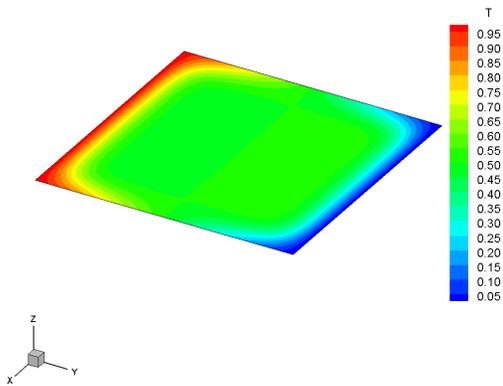
(e) Z=0.5

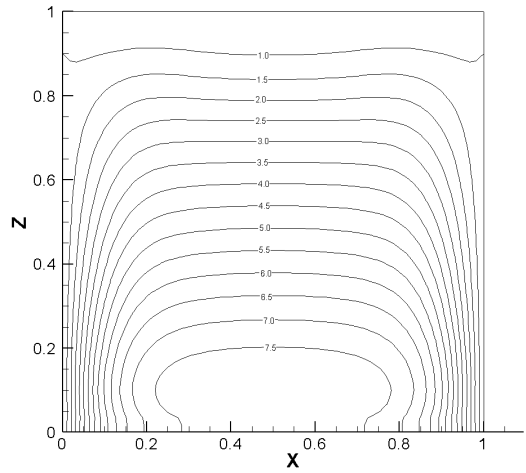
(f) Nusselt number distribution

Fig. 14

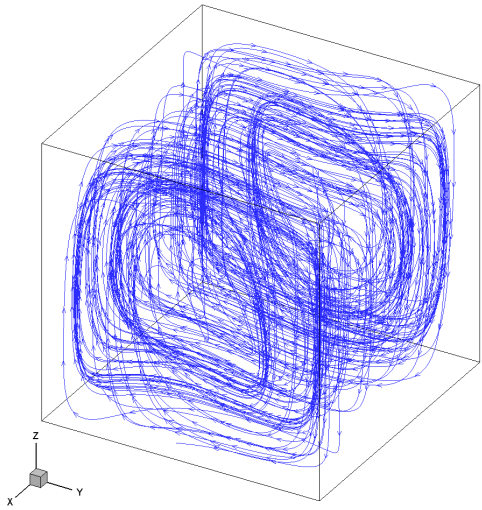
(a) 3D results

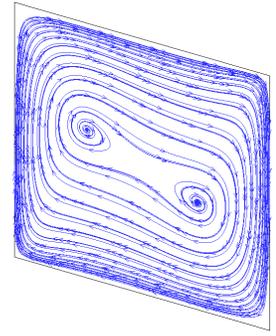
(b) X=0.5

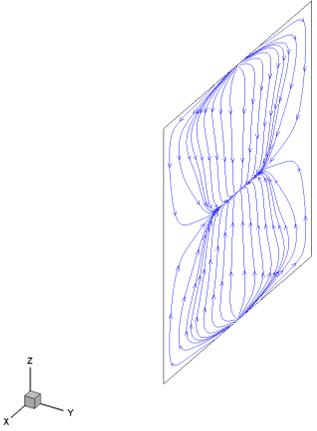
(c) Y=0.5

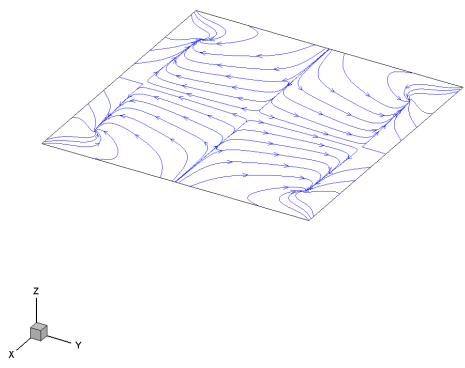
(d) Z=0.5

Fig. 15

Table 1

|  | Ref. [31] | Ref. [32] | Ref. [34] | Ref. [43] | $50\times50\times50$ | $60\times60\times60$ |
|---|---|---|---|---|---|---|
| $Ra=1\times10^4$ | 2.05 | 2.06 | 2.08 | 2.10 | 2.07 | 2.08 |
| $Ra=1\times10^5$ | 4.34 | 4.37 | 4.38 | 4.36 | 4.42 | 4.39 |

Table 2

|  | Ref. [31] | Ref. [35] | $50\times50\times50$ | $60\times60\times60$ |
|---|---|---|---|---|
| $Ra=1\times10^4$ | 2.25 | 2.30 | 2.27 | 2.27 |
| $Ra=1\times10^5$ | 4.61 | 4.67 | 4.72 | 4.69 |

Table 3

|  | Ref. [34] | $50 \times 50 \times 50$ | $60 \times 60 \times 60$ |
|---|---|---|---|
| $Ra = 1 \times 10^4$ | 3.72 | 3.66 | 3.68 |
| $Ra = 1 \times 10^5$ | 7.88 | 7.98 | 7.98 |

Table 4

|  | U | V | W |
|---|---|---|---|
| 50×50×50 | 2.20 | 16.88 | 19.23 |
| 60×60×60 | 2.19 | 16.83 | 19.20 |
| Ref. [31] | 2.16 | 16.72 | 18.98 |

Table 5

|  | U | V | W |
|---|---|---|---|
| 50×50×50 | 10.09 | 45.00 | 72.50 |
| 60×60×60 | 9.98 | 44.81 | 72.72 |
| Ref. [31] | 9.70 | 43.91 | 71.11 |

Table 6

|  | Ref. [30] | LBM |
|---|---|---|
| $Ra = 1\times10^4$ | 1.52 | 1.49 |
| $Ra = 1\times10^5$ | 3.10 | 3.06 |

Table 7

|  | $U_{max}$ | $V_{max}$ | $W_{max}$ | $\overline{Nu_{2D}}$ | $\overline{Nu_{3D}}$ | $Nu_{max}$ |
|---|---|---|---|---|---|---|
| $Ra = 1 \times 10^4$ | 3.33 | 21.18 | 22.42 | 1.75 | 1.49 | 2.68 |
| $Ra = 1 \times 10^5$ | 30.30 | 68.97 | 94.20 | 3.67 | 3.05 | 5.91 |

Table 8

|  | $U_{max}$ | $V_{max}$ | $W_{max}$ | $\overline{Nu_{2D}}$ | $\overline{Nu_{3D}}$ | $Nu_{max}$ |
|---|---|---|---|---|---|---|
| $Ra = 1 \times 10^4$ | 3.80 | 18.85 | 20.84 | 2.21 | 1.80 | 3.59 |
| $Ra = 1 \times 10^5$ | 21.73 | 59.19 | 86.26 | 4.66 | 3.94 | 7.96 |